\documentclass[twocolumn,showpacs,preprintnumbers,amsmath,amssymb,floatfix]{revtex4}

\usepackage{graphicx}
\usepackage{dcolumn}
\usepackage{bm}

\newcommand{\pom}{\tt I\! P}
\newcommand{\beq}{\begin{equation}}
\newcommand{\eeq}{\end{equation}}

\begin{document}

\title{Diffractive quarkonium production in association with a photon at the LHC}


\author{M. B. Gay Ducati$^a$, M. M. Machado$^a$}
\author{M. V. T. Machado$^b$}

\affiliation{$^a$ High Energy Physics Phenomenology Group, GFPAE,  IF-UFRGS \\
Caixa Postal 15051, CEP 91501-970, Porto Alegre, RS, Brazil}
\affiliation{$^b$ Universidade Federal do Pampa. Centro de Ci\^encias Exatas e Tecnol\'ogicas, \\
Campus de Bag\'e, Rua Carlos Barbosa. CEP 96400-970. Bag\'e, RS, Brazil}

\begin{abstract}

The rates for diffractive quarkonium plus prompt-photon associated production at the LHC are estimated.  The theoretical framework considered in the $J/\Psi$ (or $\Upsilon$) production associated with a direct photon at the hadron collider is the non-relativistic QCD (NRQCD) factorization formalism. The corresponding single diffractive cross section is computed based on the hard diffractive scattering factorization supplemented by absorptive corrections. Such processes are sensitive to the gluon content of the Pomeron at small Bjorken-x and they may also be a good place to test the different available mechanisms for quarkonium production at hadron colliders. 
\end{abstract}


\maketitle

\section{Introduction}

In recent years, a wide variety of small-$x$ and hard diffractive processes which are under intense study experimentally has been used to improve our knowledge about Quantum Chromodynamics (QCD). An outstanding process is the diffractive heavy quarkonium  production, where their large masses provide a natural hard scale that allows the application of perturbative QCD. There are several mechanisms proposed for the quarkonium production in hadron colliders \cite{Lansberg,Kramer}, as the color singlet model, the color octet model and the color evaporation model. An important feature of these perturbative QCD models is that the cross section for quarkonium  production is expressed in terms of the product of two gluon densities at large energies. In a similar way, the diffractive quarkonium production is also sensitive to the gluon content of the Pomeron at small-$x$ and may be particularly useful in studying the different mechanisms for quarkonium production.  From the experimental point of view, the heavy quarkonium production is of special significance because they have an extremely clean signature through their leptonic decay modes.

As long diffractive processes are concerned, in hadron collisions they are well described, with respect to the overall cross-sections, by Regge theory in terms of the exchange of a Pomeron  with vacuum quantum numbers \cite{Collins}. However, the nature of the Pomeron and its reaction mechanisms are not completely known. A good channel is the use of hard scattering to resolve the quark and gluon content in the Pomeron \cite{IS}. Such a parton structure is natural in a modern QCD approach to the strongly interacting Pomeron. The systematic observations of diffractive deep inelastic scattering (DDIS) at HERA have increased the knowledge about the QCD Pomeron, providing us with the diffractive distributions of singlet quarks and gluons in Pomeron as well as the diffractive structure function \cite{H1diff}. In hadronic collisions, we shall characterize an event as single diffractive if one of the colliding hadrons emits a Pomeron that scatters off the other hadron. Hard diffractive events with a large momentum transfer are also characterized by the absence of hadronic energy in certain angular regions of the final state phase space (rapidity gaps). The events fulfilling the conditions of large rapidity gaps and a highly excited hadron remnant are named single diffractive in contrast to those in which both colliding hadrons remain intact as they each emit a Pomeron. the so called central diffractive events. 

 Here, we focus on the following single diffractive processes $pp\rightarrow p+(J/\Psi+\gamma) +X$ and $pp\rightarrow p+(\Upsilon+\gamma) +X$ at the LHC energies and predict their diffractive ratios as a function of transverse momentum ($p_T$) of quarkonium state. Such  processes are interesting because the produced large $p_T$ quarkonia are relatively easy to detect through their leptonic decay modes and their transverse momenta are balanced by the associated high energy photon. As we will verify, the ratio of single diffractive production cross section to the inclusive production cross section is not strongly sensitive to the heavy quarkonium production mechanism. This channel has been recently investigated in the context of heavy ion collisions \cite{mariotto} and it can also to be compared to the previous calculations on diffractive quarkonium production \cite{qniumdiff}.

We start by the hard diffractive factorization, where the diffractive cross section is the convolution of diffractive parton distribution functions and the corresponding diffractive coefficient functions in similar way as the inclusive case. However, at high energies there are important
contributions from unitarization effects to the single-Pomeron exchange cross section. These absorptive or unitarity corrections cause the suppression of any large rapidity gap process, except elastic scattering. In the black disk limit the absorptive corrections may completely terminate those processes. This partially occurs in (anti)proton--proton collisions, where unitarity is nearly saturated at small impact parameters \cite{k3p}.  The multi-Pomeron
contributions depends, in general, on the particular
hard process and it is called survival probability factor.  At the Tevatron energy, $\sqrt s = 1.8$~TeV, the
suppression for single diffractive processes is of order 0.05--0.2~\cite{GLM,KMRsoft,BH,KKMR,Prygarin}, whereas for LHC energy, $\sqrt{s}=14$ TeV, the suppression  appears to be 0.06--0.1 ~\cite{GLM,KMRsoft,KKMR,Prygarin}.  Therefore, these corrections are quite important for the reliability of predictions for hard diffractive processes.

This Letter is organized as follows. In next section we summarize the main formulas considered to compute the diffractive ratios for the hadroproduction of $J/\psi \,(\mathrm{or}\,\Upsilon) +\gamma $  final state at the LHC energy. In the last section we present the numerical results for the inclusive e diffractive cross sections as a function of transverse momentum at the central rapidity region and give predictions for the corresponding diffractive ratios. We discuss in details the theoretical uncertainties involved in the present estimations and also compare them with previous calculations in the literature.

%
%
\section{Quarkonium plus prompt-photon associated hadroproduction}

Let us introduce the main formulas for the inclusive and single diffractive differential cross sections for the production of $J/\Psi + \gamma$ state in proton-proton collisions at high energies. For the inclusive case, we will rely on the non-relativistic QCD (NRQCD) factorization formalism and for the diffractive process the calculation are based on the Ingelman-Schlein (IS)  model for diffractive hard scattering \cite{IS}. Accordingly, we will take into account absorption effects by multiplying the diffractive cross section by a gap survival probability factor. Concerning inclusive production, as long as $J/\psi+\gamma$ is produced at small longitudinal momentum fraction,  the gluon fusion channel dominates over the $q\bar{q}$ annihilation process. The corresponding signal is the production of a $J/\psi$ and an isolated photon produced back-to-back, with their transverse momenta balanced. The leading order (LO) cross section is obtained by convoluting the partonic cross section with the parton distribution function (PDF), $g(x,\mu_F)$, in the proton, where $\mu_F$ is the factorization scale. At NLO expansion on $\alpha_s$, there are one virtual correction and three real corrections processes, as shown in Ref. \cite{Wang,Lansberg_nlo}.  In the NRQCD factorization formalism, non-perturbative aspects of quarkonium production are organized in an expansion in powers of $v$, which is the relative velocity of the $q\bar{q}$ in quarkonia. The partonic level subprocesses for associated $J/\psi+\gamma$ production are composed of the gluon fusion subprocesses, which are
\begin{eqnarray}\nonumber
g + g & \rightarrow & \gamma+(c\bar{c})\left [ ^{3}S^{1}_{1}, ^{3}S^{8}_{1} \right ] \, , \\
g + g & \rightarrow & \gamma+(c\bar{c})\left [ ^{1}S^{8}_{0}, ^{3}P^{8}_{J} \right ] \, .
\label{equacao1}
\end{eqnarray}
Quarks initiated subprocesses are strongly suppresses and then neglected further.

In order to obtain the transverse momentum ($p_T$) distribution for the process $g+g\rightarrow J/\psi + \gamma + g$, we express the differential cross section as
\begin{eqnarray}
\frac{d^2\sigma_{\mathrm{inc}}}{dydp_T}=\int dx_1g_p(x_1,\mu_F^2) g_p(x_2,\mu_F^2)\frac{4x_1x_2p_T}{2x_1-\bar{x}_Te^y}\frac{d\hat{\sigma}}{d\hat{t}}\,,
\label{diffcs}
\end{eqnarray}
where we have defined $\bar{x}_T=2m_T/\sqrt{s}$, with $\sqrt{s}$ being the center of mass energy of the $pp$ system and $m_T=\sqrt{p_T^2+m_{\psi}^2}$ being the transverse mass of outgoing $J/\psi$. The gluon distribution, $g_{p}(x,Q^2)$, in the proton is evaluated at factorization scale $\mu_F$. The common transverse momentum of the outgoing particles is $p_T$ and $y$ is the rapidity of outgoing $J/\psi$ having mass $m_{\psi}$. The variables $x_1$ and $x_2$ are the momentum fractions of the partons, where $M^2/s\leq x_1<1$ ($M$ is the invariant mass of $J/\psi+\gamma$ system) and $x_2$ can be written in terms of other variables as
\begin{eqnarray}
x_2=\frac{x_1\bar{x}_Te^{-y}-2\tau}{2x_1-\bar{x}_Te^y}, \,\,\,\,\mathrm{with} \,\,\,\,\tau=\frac{m_{\psi}^2}{s}.
\end{eqnarray}

In the NRQCD formalism, the cross section for the production of a quarkonium state $H$ is written as $\sigma (H)=\sum_n c_n\langle 0|O_n^H|0 \rangle$, where the short-distance coefficients $c_n$ are computable in perturbation theory. The $\langle 0|O_n^H|0 \rangle$ are matrix elements of NRQCD operators of the form
\begin{equation}\label{Matrix}
\langle 0|O_n^H|0 \rangle = \sum_X \sum_{\lambda} \langle 0|\kappa_n^{\dagger}|H(\lambda) + X\rangle \langle H(\lambda) + X |\kappa_n|0 \rangle.
\end{equation}
The $\kappa_n$ is a bilinear in heavy quark fields
which creates a $Q \overline{Q}$ pair in a state with definite color and angular momentum quantum numbers. Hereafter, we will use a shorthand notation in which the matrix elements are given as $\langle O^H_{(1,8)}(^{2S+1}L_J) \rangle$. The angular momentum quantum numbers of the $Q\overline{Q}$
produced in the short-distance process are given in standard spectroscopic notation, and the subscript refers to the color configuration of the $Q\overline{Q}$: $1$ for a color singlet and $8$ for a color octet. For example, the parton level differential cross sections relevant for hadroproduction
of $J/\psi + \gamma$ for the color-singlet contributions is given by \cite{Mehen,Kim2}:
\begin{eqnarray}\label{GGPsiGa}
{d\hat{\sigma}_{\mathrm{sing}} \over d\hat{t}}  =  \sigma_0 \left[ {16\over 27} \left( {\hat{s}^2 s_1^2 + \hat{t}^2 t_1^2 + \hat{u}^2 u_1^2 \over s_1^2 t_1^2 u_1^2} \right) \langle O_1^{J/\psi}(^3S_1)\rangle \right],
\label{Jpsixsection}\end{eqnarray}
and the full expressions for the color octet contributions can be found, for instance, on Refs. \cite{Mehen,Kim2}.

In Eq. (\ref{GGPsiGa}) we have defined $\sigma_0=\pi^2 e_c^2 \alpha \alpha_s^2 m_c/ \hat{s}^2$ (with charm quark mass $m_c=1.5$ GeV) and $s_1 = \hat{s} - 4 m_c^2$, $t_1 = \hat{t} - 4 m_c^2$, and  $u_1 = \hat{u} - 4 m_c^2$. In these formulas, $\hat{s}$, $\hat{t}$, and $\hat{u}$ are the Mandelstam variables, which can be written as:
\begin{eqnarray}
\hat{s}=x_1x_2s,\hspace{0.1cm} \hat{t}=m_{\psi}^2-x_2\sqrt{s}m_{T}e^y, \hspace{0.1cm} \hat{u}=m_{\psi}^2-x_1\sqrt{s}m_{T}e^{-y}, \nonumber
\end{eqnarray}

In our numerical calculations the one loop expression for the running coupling, $\alpha_s(\mu_R)$, with $\Lambda_{QCD}=0.2$ GeV and $n_f=4$ is considered. The (renormalization and factorization) scale for the strong coupling and for the evaluation of PDFs is $\mu_F^2=\mu_R^2=(p_T^2+m_{\psi}^2)/4$. For numerical values of the NRQCD matrix elements we have used those from Ref. \cite{Matrix}, which are (units of $GeV^3$): $\langle O_1^{J/\psi}(^3S_1)\rangle=1.16$, $\langle O_8^{J/\psi}(^3S_1)\rangle=1.19\times 10^{-2}$, $\langle O_8^{J/\psi}(^1S_0)\rangle=\langle O_8^{J/\psi}(^1P_0)\rangle/m_c^2=0.01$. We have checked that using another set of color octet matrix elements, taken from \cite{Kramer:2001}, our results do not change considerably. The production of $\Upsilon(1S)+\gamma$ can be obtained from the expressions (\ref{diffcs}-\ref{GGPsiGa}) above, by replacing the charm mass and charge by the bottom ones ($m_b=4.7$ GeV, $e_b=-1/3$), the $J/\psi$ mass by the $\Upsilon (1S)$ mass, and by using the corresponding matrix elements. We use the values taken from \cite{Braaten:2001}, namely (units of $GeV^3$): $\langle O_1^{\Upsilon}(^3S_1)\rangle=10.9$, $\langle O_8^{\Upsilon}(^3S_1)\rangle=0.02$, $\langle O_8^{\Upsilon}(^1S_0)\rangle=0.136$ and $\langle O_8^{\Upsilon}(^1P_0)\rangle=0$. 

 Lets move now to the hard diffractive cross section. The associated $J/\psi+\gamma$ single diffractive production at large $p_{T}$ consists of three steps: first a hard Pomeron is emitted from one of the protons in a small squared four-momentum transfer $|t|$. After that, partons from the Pomeron interact with partons from the other hadron. Finally, $J/\psi+\gamma$ (or $\Upsilon + \gamma$) are produced in the final state, from the point-like $Q\bar{Q}$ by the soft gluon radiation.
Considering the $p_{T}$ distribution and based on the IS model for diffractive hard scattering, the differential cross section can be expressed as
\begin{eqnarray}
\frac{d^2\sigma_{\mathrm{SD}}}{dydp_T}& = & \int_{x_{\pom}^{min}}^{x_{\pom}^{max}}dx_{\pom} \int_{\frac{M^2}{sx_{\pom}}}^{1}dx_1\int_{-1}^0dt\,f_{\pom/p}(x_{\pom},t) \nonumber \\
& \times & g_{\pom}(x_1,\mu_F^2) g_p(x_2,\mu_F^2)\frac{4x_1x_{\pom}x_2p_T}{2x_1x_{\pom}-\bar{x}_Te^y}\frac{d\hat{\sigma}}{d\hat{t}}\,,
\label{diffsd}
\end{eqnarray}
where $x_{\pom}$ is the momentum fraction of the proton carried by the Pomeron, $t$ is the squared of the proton's four-momentum transfer, $x_{1}$ is the momentum fraction of the partons inner the Pomeron and $x_{2}$ is the momentum of the partons inner the anti-proton. Here, $f_{\pom /p}(x_{\pom},t)$ is the Pomeron flux factor, which can be written as $f_{\pom /p}(x_{\pom},t) \propto x_{\pom}^{1-2\alpha(t)}F^{2}(t)$. In the following calculation, we use the Pomeron factor flux considered by the H1 Collaboration \cite{H1diff} analysis of the diffractive structure function in diffractive DIS process. The variable $x_2$ and the corresponding Mandelstam variable read now as 
\begin{eqnarray}
x_2 & = &\frac{x_1x_{\pom}\bar{x}_Te^{-y}-2\tau}{2x_1x_{\pom}-\bar{x}_Te^y},\\
\hat{s} & = & x_1x_2x_{\pom}s,\hspace{0.5cm} \hat{t}=m_{\psi}^2-x_2\sqrt{s}m_{T}e^y, \\
\hat{u} & = & m_{\psi}^2-x_1x_{\pom}\sqrt{s}m_{T}e^{-y}.
\end{eqnarray}

To calculate the hard diffractive cross sections, Eq. (\ref{diffsd}), we consider a standard Pomeron flux from Regge phenomenology and which is constrained from the experimental analysis of the diffractive structure function \cite{H1diff}, where $x_{\pom}$ dependence is parametrized as
\begin{eqnarray}
f_{\pom/p}(x_{\pom,t})=A_{\pom}\cdot \frac{e^{B_{I\pom}t}}{x^{2\alpha_{\pom}(t)-1}_{\pom}}\, ,
\label{equacao13}
\end{eqnarray}
with the Pomeron trajectory assumed to be linear, $\alpha_{\pom}(t)=\alpha_{\pom}(0)+\alpha^{\prime}_{\pom}t$ and $B_{\pom}$, $\alpha^{\prime}_{\pom}$ and their uncertainties are obtained from fits to H1 FPS data \cite{H1FPS}. $A_{\pom}$ is the normalization parameter, choosed such that $x_{\pom}\cdot \int^{t_{min}}_{t_{cut}}f_{\pom/p}dt=1$ at $x_{\pom}=0.003$, where $|t_{min}|\approx m^{2}_{p}x^{2}_{\pom}/(1-x_{\pom})$ is the minimum kinematically accessible value of $|t|$, $m_{p}$ is the proton mass and $|t_{cut}|=1.0$ GeV$^{2}$ is the limit of the measurement. For the diffractive gluon distribution in the Pomeron, $g_{\pom}(x_1,\mu_F^2)$, we will consider the diffractive PDFs obtained by the H1 Collaboration at DESY-HERA \cite{H1diff}, where the Pomeron structure function has been modeled in terms of a light flavor singlet distribution $\Sigma(x)$, i. e., the $u$, $d$ and $s$ quarks with their respective antiquarks. Also, it has a gluon distribution $g(z)$, with $z$ being the longitudinal momentum fraction of the parton entering the hard subprocess with respect to the diffractive exchange. The gluon density is a simple constant at the starting scale for evolution, which was chosen to be $Q^{2}_{0}=2.5$ GeV$^{2}$. In our numerical calculations, we will use the following cuts for the integration over $x_{\pom}$,  $x_{\pom}^{min}\leq x_{\pom}\leq 0.05$, where $x_{\pom}^{min}=\frac{\bar{x}_Te^{y}-2\tau}{\bar{x}_Te^{-y}-2},$.

\begin{figure}[t]
\includegraphics[scale=0.35]{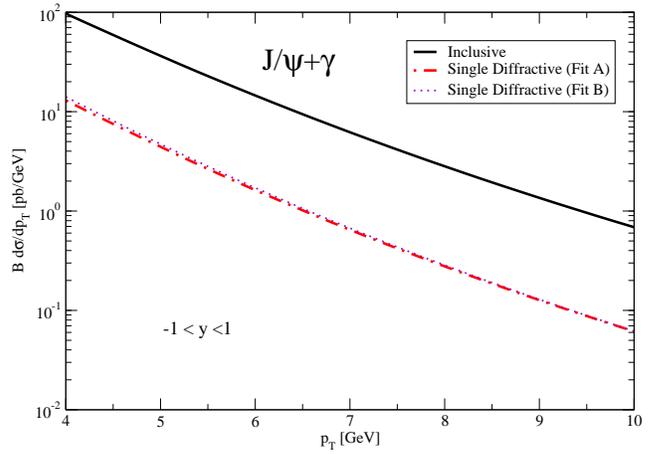}
\caption{The inclusive (upper solid line) and single diffractive (bottom lines) cross section for  $J/\Psi+\gamma$ hadroproduction as a function of $p_{T}$ at the LHC. The diffractive cross section  is computed for two sets of diffractive gluon distribution (FIT A and FIT B). Only central rapidity region, $|y | \leq 1$, is considered.}
\label{fig:1}
\end{figure}

As a final ingredient for the single diffractive cross section, we also consider the suppression of the hard diffractive cross section by multiple-Pomeron scattering effects (absorptive corrections). This is taken into account through a gap survival probability $<|S|^{2}>$, which can be described in terms of screening or absorptive corrections \cite{Bj}. Let ${\cal{A}}(s,b)$ be the amplitude of the particular diffractive process of interest and considering it in the impact parameter space $b$, the probability that there is no extra inelastic interaction is
\begin{eqnarray}
<|S|^{2}>=\frac{\int d^{2}b|{\cal{A}}(s,b)|^{2}exp\left [-\Omega(s,b)\right ] }{\int d^{2}b |{\cal{A}}(s,b)|^{2}}\, ,
\label{equacao14}
\end{eqnarray}
where $\Omega$ is the opacity (or optical density) of the interaction of the incoming hadrons. There are intense theoretical investigation on this subject in last years. We quote Ref. \cite{Prygarin} for a nice discussion and comparison of theoretical estimations for the gap survival probabilities. We notice that it is the main theoretical uncertainty in the present calculation of diffractive ratios. As a baseline value, we follow  Ref. \cite{KKMR}, which considers a two-channel eikonal model that embodies pion-loop insertions in the Pomeron trajectory and high mass diffractive dissociation. For LHC energy on $pp$ collisions, one has $<|S|^{2}>=0.06$. In single channel eikonal models, this factor can reach up to $0.081-0.086$ as discussed in Ref. \cite{Prygarin}.  Concerning the model dependence, the single channel eikonal model considers only elastic rescatterings, whereas the multi channel one takes into account also inelastic diffractive intermediate re-scatterings.  The available experimental observables which can be compared to the theoretical predictions of the survival probability factor are the hard LRG di-jets data obtained in the Tevatron and HERA \cite{Prygarin,KKMR}  as well as diffractive hadroproduction of heavy bosons ($W^{\pm}$ and $Z^0$) in the Tevatron \cite{GDMM}.

\subsection{Results and comments}

\begin{figure}[t]
\includegraphics[scale=0.35]{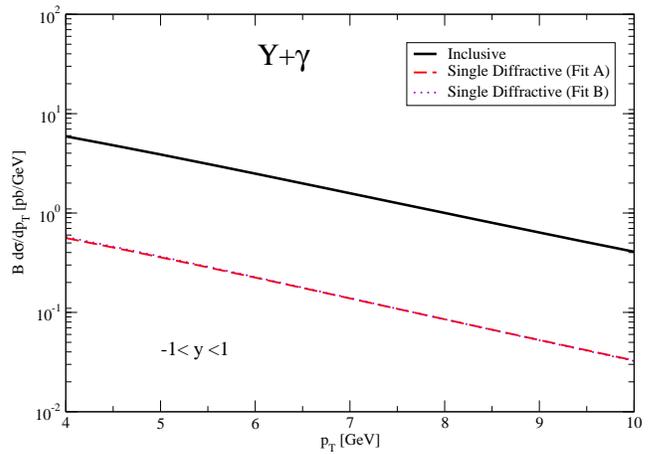} 
\caption{The inclusive (upper solid line) and single diffractive (bottom lines) cross section for  $\Upsilon\,(1S)+\gamma$ hadroproduction as a function of $p_{T}$ at the LHC. The diffractive cross section  is computed for two sets of diffractive gluon distribution (FIT A and FIT B). Only central rapidity region, $|y | \leq 1$, is considered.}
\label{fig:2}
\end{figure}

Let us now present the estimates for the quarkonium plus prompt-photon associated diffractive hadroproduction for the LHC energy. In Fig. \ref{fig:1} we present the numerical calculations for the inclusive cross section (upper solid line) for $J/\Psi + \gamma$ production as a function of transverse momentum, given by Eq. (\ref{diffcs}), using the MRST gluon PDF \cite{MRST}. The numerical results include both the color singlet and color octet contributions to the production process. The corresponding branching ratio into dileptons is taken into account and we concentrate on the central rapidity region, imposing the kinematic  cut $|y|\leq 1$. The order of magnitude is not small, giving units of nanobars in the fully integrated case. Notice that the absolute value of inclusive cross section is strongly dependent on the quark mass, NRQCD matrix elements and on the factorization scale and it is know \cite{Wang,Lansberg_nlo} that the NLO corrections strongly enhances the cross section at large $p_T$. This theoretical uncertainty is minimized when we compute the corresponding diffractive ratio. The single diffractive cross section (without absorption corrections) is shown in the bottom lines. We present the estimate using two different sets of the diffractive gluon distribution function determined experimentally. The result for FIT A parameterization is represented by the dot-dashed curve, whereas the FIT B calculation is given by the thin dashed curve. It is verified that the absolute value of diffractive cross section is weakly sensitive to the uncertainties on the diffractive gluon PDF, being more pronounced in the charmonium case and on the small $p_T$ region.  We perform a similar analysis for production of the bottomonium state $\Upsilon (1S)$.  In order to do so, in Fig. \ref{fig:2} the inclusive cross section and the single  diffractive cross section for production of  final state  are presented as a function of $p_T$. The label for the curves follows those already considered in Fig. \ref{fig:1}. The cross sections are about one order of magnitude smaller than the charmonium case. This is understood from the dependences on  quark charge and mass $\sigma \propto e_Q^2/m_Q^2$, which gives roughly speaking $\sigma_{\psi}/\sigma_{\Upsilon}\propto (e_c^2/e_b^2)[m_b^2/m_c^2]=4  [m_b^2/m_c^2]$. As already mentioned, the single diffractive cross section is almost insensitive to the diffractive gluon PDF in the range on $p_T$ presented here. The situation would be different for the central diffractive case, i.e. the double Pomeron exchange (DPE)  process, which is dependent on the product of diffractive gluon distribution on both Pomerons. Notice that our numerical results are somewhat consistent with those presented in Ref. \cite{Xu}, where the $J/\Psi + \gamma$ cross section is computed in the same framework and a renormalized Pomeron flux is considered for the investigation of single diffractive cross section. Our estimate for the Upsilon  (diffractive) case has been not addressed before in literature. 

Our main results are presented in Table I, where the prediction for the diffractive ratio as a function of $p_T$ is shown (using FIT A for the diffractive PDFs). As a sample of numerical results for the inclusive and single diffractive cross section, we also give the estimates for charmonium and bottomonium plus prompt-photon associated hadroproduction in the LHC regime. The transverse momenta are presented in the range $4\leq p_T\leq 10$ GeV for central rapidity region $|y|<1$. It was verified that when considering the forward rapidity region (for instance, $|y|\geq 2$) a smaller diffractive ratio is found. This is consistent with previous calculation, presented in Ref. \cite{Xu}. However, we have found a slightly large diffractive ratio in comparison to that work. The reason can be twofold: for the inclusive case, we are using a different factorization scale ($\mu_F=E_T$  in Ref. \cite{Xu}) and in the diffractive case we considered the eikonal-inspired absorptive corrections against renormalized Pomeron flux \cite{Goulianos} taken in Ref. \cite{Xu} (the authors also neglected the $Q^2$ evolution in the gluon density of the Pomeron). This last fact could also to explain the $p_T$ dependence appearing in our estimates for the diffractive ratios, whereas in Ref. \cite{Xu} such a dependence is weaker.

As a summary, the theoretical predictions for inclusive and single diffractive $J/\psi+\gamma$ and $\Upsilon + \gamma$ production has been done for the LHC energy in $pp$ collisions. The estimates for the differential cross sections as a function of quarkonium transverse momentum are presented, discussing the main theoretical uncertainties. The corresponding diffractive ratio is computed using hard diffractive factorization and absorptive corrections. It is verified that the ratios are less dependent on the heavy quarkonium production mechanism and quite sensitive to the absolute value of absorptive corrections.  We found that at the LHC, the diffractive ratio for $J/\psi+\gamma$ state is about $R_{\mathrm{SD}}=0.8-0.5 \,\%$ in the interval  $4\leq p_T\leq 10$ GeV and $R_{\mathrm{SD}}=0.6-0.4 \,\%$ in same interval for the $\Upsilon+\gamma$ state.


\begin{table}
\caption{\label{tabela} Inclusive and single diffractive cross sections in units of picobarns for the interval $ 4 \leq p_{T} \leq 10$ GeV for central rapidity region (integrated over $|y|\leq 1$). The corresponding diffractive ratios, $R_{\mathrm{SD}}$,  are presented considering the KKMR  value for the absorptive corrections at the LHC energy, $<|S|^{2}>=0.06$. }
\begin{ruledtabular}
\begin{tabular}{l|c|c|c|c|c|c|c}
    $p_{T}$ [GeV]    &   $4$	&    $5$   &   $6$    &    $7$    &    $8$   &    $9$	&   $10$        \\\hline
$\frac{d\sigma_{\mathrm{inc}}}{dp_T}$ ($J/\Psi$)     &	$97.04$	&  $36.46$ & $14.54$  &  $6.21$   &  $2.82$  &  $1.36$  &  $0.68$   	\\\hline
$\frac{d\sigma_{\mathrm{SD}}}{dp_T}$ ($J/\Psi$)  &	$0.78$	&  $0.26$  & $0.10$   &  $0.04$   &  $0.017$ &  $0.008$ &  $0.0036$ 	\\\hline
$R_{\mathrm{SD}}$ [\%]         & $0.8$ 	&  $0.71$  & $0.69$   &  $0.64$   &  $0.6$   &  $0.59$  &  $0.53$   	\\ \hline
 $\frac{d\sigma_{\mathrm{inc}}}{dp_T}$ ($\Upsilon$)      &	$5.91$	&  $3.88$  & $2.49$  &  $1.58$   &  $1.00$  &  $0.64$  &  $0.41$      \\\hline
$\frac{d\sigma_{\mathrm{SD}}}{dp_T}$ ($\Upsilon$)    &	$0.036$	&  $0.022$ & $0.013$ & $0.008$  &  $0.0054$ &  $0.003$ &  $0.0018$    \\\hline
$R_{\mathrm{SD}}$ [\%]      &   $0.6$ 	&  $0.56$  & $0.53$  &  $0.51$   &  $0.54$   &  $0.47$  &  $0.44$     
\end{tabular}
\end{ruledtabular}
\end{table}


%
%


\section*{Acknowledgments}

This work was supported by CNPq and FAPERGS, Brazil.


%
%


\end{document}